\documentclass[useAMS,usenatbib]{mn2e}

\usepackage{epsfig}
\usepackage{amsmath}
\usepackage{amssymb}
\usepackage{euscript}

\newcommand{\perbeam}{\,beam$^{-1}$}

\newcommand{\dd}{{\rm d}}

\title[Low-level activity in Cygnus X-3]{Opacity effects and
  shock-in-jet modelling of low-level activity in Cygnus X-3}
  \author[J.C.A.~Miller-Jones et al.]  {James
  C.A.~Miller-Jones,$^{1}$\thanks{E-mail: jmiller@nrao.edu} Michael
  P.~Rupen,$^2$ Marc T\"urler,$^{3,4}$ \and Elina J.~Lindfors,$^{5,6}$
  Katherine M.~Blundell,$^7$ and Guy G.~Pooley.$^{8}$\\ $^1$Jansky
  Fellow, National Radio Astronomy Observatory, 520 Edgemont Road,
  Charlottesville, VA 22903, USA\\ $^2$NRAO, Array Operations Center,
  1003 Lopezville Road, Socorro, NM 87801, USA\\ $^3$Geneva
  Observatory, University of Geneva, Ch.\ des Maillettes 51, 1290
  Sauverny, Switzerland\\ $^4$ISDC Data Centre for Astrophysics, Ch.\
  d'Ecogia 16, 1290, Versoix, Switzerland\\ $^5$Tuorla Observatory,
  Department of Physics and Astronomy, University of Turku, 21500
  Piikki\"o, Finland\\ $^6$Mets\"ahovi Radio Observatory, Helsinki
  University of Technology, 02540 Kylm\"al\"a, Finland\\
  $^7$University of Oxford, Astrophysics, Keble Road, Oxford, OX1 3RH,
  UK\\ $^{8}$Astrophysics, Cavendish Laboratory, J.J.~Thomson Avenue,
  Cambridge CB3 0HE }

\begin{document}

\date{Draft of \today}

\pagerange{\pageref{firstpage}--\pageref{lastpage}} \pubyear{2008}

\maketitle

\label{firstpage}

\begin{abstract}
We present simultaneous dual-frequency radio observations of Cygnus
X-3 during a phase of low-level activity.  We constrain the minimum
variability timescale to be 20\,min at 43\,GHz and 30\,min at 15\,GHz,
implying source sizes of 2--4\,AU.  We detect polarized emission at a
level of a few per cent at 43\,GHz which varies with the total
intensity.  The delay of $\sim10$\,min between the peaks of the flares
at the two frequencies is seen to decrease with time, and we find that
synchrotron self-absorption and free-free absorption by entrained
thermal material play a larger role in determining the opacity than
absorption in the stellar wind of the companion.  A shock-in-jet model
gives a good fit to the lightcurves at all frequencies, demonstrating
that this mechanism, which has previously been used to explain the
brighter, longer-lived giant outbursts in this source, is also
applicable to these low-level flaring events.  Assembling the data
from outbursts spanning over two orders of magnitude in flux density
shows evidence for a strong correlation between the peak brightness of
an event, and the timescale and frequency at which this is attained.
Brighter flares evolve on longer timescales and peak at lower
frequencies.  Analysis of the fitted model parameters suggests that
brighter outbursts are due to shocks forming further downstream in the
jet, with an increased electron normalisation and magnetic field
strength both playing a role in setting the strength of the outburst.
\end{abstract}

\begin{keywords}
X-rays: binaries -- shock waves -- radio continuum: stars -- stars:
Cygnus X-3 -- ISM: jets and outflows
\end{keywords}

\section{Introduction}
Cygnus X-3 is one of the few persistently bright radio-emitting X-ray
binary systems in our Galaxy.  It exhibits resolved relativistic jets
during its occasional giant outbursts, which have been directly imaged
on milliarcsecond scales \citep{Mio01,Mil04,Tud07} with very long
baseline interferometry (VLBI).  While the giant outbursts are the
most spectacular and well-studied type of radio behaviour, the system
remains active at a lower level for most of the time.  \citet{Wal94}
identified three different types of variability; normal quiescent
emission (60--140\,mJy, variable on a timescale of months), periods of
frequent small flaring ($<1$\,Jy), and the major ($>10$\,Jy) or
intermediate ($>1$\,Jy) flares, with their associated pre-flare quench
periods.  These classifications were based on dual-frequency data from
the Green Bank Interferometer, with 10-min observations made 3--5
times per day.  \citet{Wal96} deduced from the higher-than-expected
scatter in the quiescent flux densities that irregular flux variations
continue to be present in this low-level state, particularly at the
higher frequencies, and data at higher time resolution has shown that
the so-called ``quiescent'' state is in fact variable on much shorter
timescales (Molnar, Reid \& Grindlay 1984, 1985, 1988), of order tens
of minutes.  Far from being quiescent, the source is continuously
active at a low level, aside from the pre-flare periods of truly
quenched emission.

Newell, Garrett \& Spencer (1998) observed some minor flaring episodes
(peak fluxes $\sim300$\,mJy) with the Very Long Baseline Array (VLBA)
and claimed evidence for superluminal expansion and contraction of the
source.  No other high-resolution observations have addressed the
nature of the source during the low-level, active state in which it
spends the majority of its time.  The resolved nature of the source in
the observations of \citet{New98} demonstrated that the jets are still
active even outside the major flaring events.  Direct evidence for jet
activity outside flaring events is also seen in GRS\,1915+105, in
which an AU-scale nuclear jet was resolved with the VLBA (Dhawan,
Mirabel \& Rodr\'\i guez, 2000).  At the resolution of the
images, the jet was observed to be continuous rather than knotty,
with a length which varied with observing wavelength as
$\sim10\lambda_{\rm cm}$\,AU.  The lightcurves of the observations
confirmed that this was an optical depth effect.  We see emission from
the $\tau=1$ surface, which is further out at lower frequencies.  This
then results in a time delay between flux variations at different
frequencies, as any injection of relativistic plasma at the jet base
takes time to propagate out to the point at which the jet is optically
thin at the observing frequency \citep[as seen in GRS\,1915+105
by][]{Mir98}.  This also gives rise to the smoothing out of the
variability at lower frequencies, since the observed emission is a
convolution over a larger region.  Measurements of the time delay
between different observing frequencies can thus help constrain the
size scale of the jets.

Being optically thick, such self-absorbed jets tend to show very low
levels of linear polarization.  \citet{Cor00} detected linear
polarization at a level of $\sim2$ per cent in the jets of GX\,339-4
while the source was in the low/hard X-ray state (believed to be
associated with steady, low-level jets).  The position angle of the
electric field vector remained constant over more than two years,
indicating a favoured axis in the system, believed to be aligned with
the axis of the compact jets.  Transient jets, on the other hand,
regularly show significantly higher levels of polarized emission,
arising from optically thin ejecta \citep[e.g.][]{Fen99,Han00,Bro07},
which may vary in amplitude and position angle with time, as different
components dominate the flux density at different times, or if the
source moves out from behind a screen of ionised electrons which
rotate the position angle via Faraday rotation.

The relativistically-moving transient ejecta observed during X-ray
binary outbursts \citep[e.g.][]{Mir94,Fen99,Tin95,Hje95} have
previously been modelled either as discrete knots of
adiabatically-expanding plasma \citep[e.g.][]{van66,Hje88,Ato99}, or
as internal shocks within a pre-existing steady flow (e.g., Fender,
Belloni \& Gallo, 2004).  \citet{Ato99} found that a discrete plasmon
model requires continuous replenishment of relativistic particles as
well as radiative, adiabatic, and energy-dependent escape losses.
Kaiser, Sunyaev \& Spruit (2000) argued that such energy-dependent
escape losses were inconsistent with continuous particle replenishment
via shock acceleration, and adapted the internal shock model
originally derived for AGN jets \citep{Ree78,Mar85} to X-ray binary
systems.  \citet{Fen04} integrated the internal shock scenario into a
model explaining the disc-jet coupling over the entire duty cycle of
an X-ray binary system.  In this scenario, as a source moves from a
hard to a soft X-ray state, the bulk Lorentz factor of the
pre-existing steady jet increases, giving rise to shocks where the
highly-relativistic plasma catches up with pre-existing slower-moving
material downstream in the jets.  In this work, we focus on the
shock-in-jet model, since internal shocks form a natural explanation
for the outbursts in the current disc-jet coupling picture
\citep{Fen04}, and since shocks provide a mechanism for the continuous
replenishment of relativistic particles found to be required in a
plasmon model \citep{Ato99}.  While detailed model fitting for the
plasmon scenario is deemed to be beyond the scope of this paper, where
relevant, we will provide brief, generic outlines of how the two
classes of model differ.

In Section \ref{sec:obs}, we present dual-frequency radio lightcurves
of Cygnus X-3, taken when the source was in its normal low-level
active state, in order to constrain the polarization of the source and
investigate the opacity effects in the jet (Section
\ref{sec:opacity}).  In Section \ref{sec:modelling}, we fit the
multifrequency lightcurves with a shock-in-jet model, in order to draw
comparisons with the properties of the giant outbursts.

\section{Observations}
\label{sec:obs}
Cygnus X-3 was observed for 8\,hours on 2002 January 25 (MJD\,52299)
with the Very Large Array (VLA) in its most extended A-configuration,
under program code AR\,458.  The VLA was split into two subarrays, and
the source was observed simultaneously at 14.940\,GHz in one subarray
and at 43.340\,GHz in the other.  Occasional snapshots at 1.425,
4.860, 8.460 and 22.460\,GHz were also carried out in order to
characterise the overall radio spectrum.  Observations at each
frequency were made in the standard VLA continuum mode (dual circular
polarization in two 50-MHz bands, with full polarization products
being recorded).  The primary calibrator used was 3C\,286
(J\,1331+305), and the secondary calibrator was J\,2007+404.  The
absolute flux density of 3C\,286 was set using the coefficients of
\citet{Baa77}.  For the majority of the observing run, calibrator
observations of duration $\sim 1$\,min were interspersed with
10--20-min scans on Cygnus X-3.  For 25\,min in the middle of the run
however, fast-switching was used, slewing back and forth between the
target and the calibrator with a cycle time of 150\,s, spending 100\,s
on the target and 40\,s on the calibrator.  This approach was designed
to reduce tropospheric phase variation, allowing for
diffraction-limited imaging, which on the long baselines at the high
observing frequencies used might not otherwise have been possible.
The data were reduced using standard procedures within the {\sc aips}
data reduction package \citep{Gre03}.  Primary referenced pointing
(pointing up on a calibrator at a lower frequency to derive the
relevant pointing offsets for higher frequencies where the offsets may
be a significant fraction of the primary beam) was used at both 43 and
15\,GHz, with the times of the pointing scans shown in
Fig.\,\ref{fig:variability}.  An integration time of 3.3\,s was used
throughout the observing run (except for the pointing scans, where
10-s integrations were mandatory) to allow for high time-resolution
data editing.

The weather was good throughout the observations, with clear skies and
wind speeds of $<3$\,km\,s$^{-1}$, so the pointing accuracy should not
have been significantly affected by wind loading of the antennas, and
the atmospheric opacity should have been fairly stable.  In order to
accurately determine the flux density scale, gain curves measured by
NRAO staff in 2001 November were used to correct for gravitational
deformation of the antennas with changing elevation.  Opacity
corrections were made using weather (WX) tables and the fitted opacity
curves of \citet{But02}.

Cygnus X-3 was also observed at 15\,GHz by the Ryle Telescope, as part
of an ongoing monitoring programme \citep{Poo97}.  The Ryle
observations overlapped our own VLA observations for 1.6\,h at the
beginning of the VLA run.  A systematic discrepancy in the absolute
flux scales at the two instruments was found, with the VLA
measurements being $\sim25$ per cent brighter, exactly as noted by
\citet{Lin07}.  Scaling the Ryle data points by this factor brought
the two lightcurves into agreement.  The Ryle Telescope observes a
single linear polarization (Stokes $I+Q$), as opposed to the dual
circular polarization measured by the VLA, although the results in
Section \ref{sec:polarization} show that this difference is not
significant for Cygnus X-3, and is unlikely to be the explanation for
the discrepancy.

\subsection{Variability}
\label{sec:lcs}
\begin{figure}
\epsfig{figure=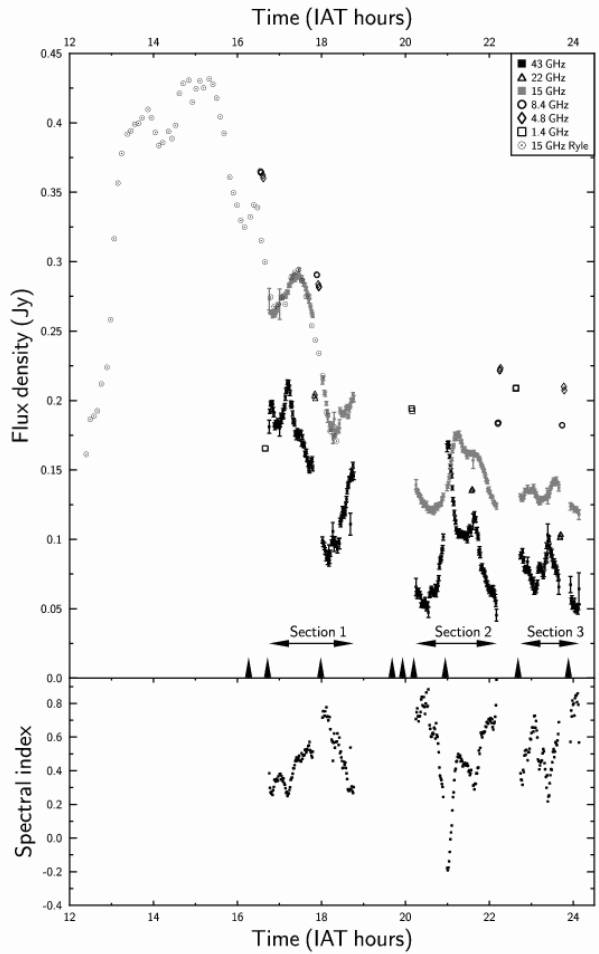,width=8cm,angle=0,clip=}
\caption[VLA lightcurves during minor flaring of 2002 January
      25]{\label{fig:variability}Top panel shows lightcurves of the
      Ryle observations at 15\,GHz (dotted circles) and the VLA
      observations at 43\,GHz (black) and 15\,GHz (grey).  Also shown
      are measurements at 22\,GHz (hollow triangles), 8.4\,GHz (hollow
      circles), 4.8\,GHz (hollow diamonds) and 1.4\,GHz (hollow
      squares).  0\,h IAT corresponds to MJD 52299.0.  Vertical arrows
      on the time axis denote times of referenced pointing observations.
      Bottom panel shows the instantaneous spectral index between 43
      and 15\,GHz, $\alpha^{15}_{43}$.}
\end{figure}

The lightcurves of the observations and the spectral index $\alpha$
(where $S_{\nu} \propto \nu^{-\alpha}$) between 43 and 15\,GHz are
shown in Fig.\,\ref{fig:variability}.  It was found that the flux
density of J\,2007+404 appeared to vary slightly during the
observations.  This can almost certainly be attributed to the pointing
drifting off, since the trend was for the derived flux density to
decline between pointing observations, especially at the higher of the
two frequencies.  In order to compensate for this effect, the
J\,2007+404 points were all corrected to a mean flux density of
1.53\,Jy at 43\,GHz and 2.83\,Jy at 15\,GHz, using a multiplicative
scaling factor.  The Cygnus X-3 flux densities were then corrected by
linearly interpolating between the gains thus derived for the
secondary calibrator.

The source is in the optically thin decay phase following the flare at
$\sim 15$\,h, with smaller flares superimposed on this steady decrease
in flux density. The amplitude of these events is greater at 43\,GHz
than at 15\,GHz, and they peak 10--15\,min earlier at the higher
frequency, reminiscent of the behaviour observed in GRS\,1915+105
\citep[e.g.][]{Poo97,Mir98,Fen02}.

\subsection{Polarization}
\label{sec:polarization}
With the wide parallactic angle coverage of an 8-hour observation, it
was possible to solve for the ``D-terms'' (the leakage of right
circular polarization signal into left circular polarization feeds and
vice versa) using the calibrator J\,2007+404.  Since 3C\,286 was
resolved by our combination of frequency and array configuration, we
used J\,2136+006 as a position angle calibrator (assuming a position
angle of 26.9\degr\ at 43\,GHz and 70.0\degr\ at 15\,GHz, derived from
monitoring data taken the following day by NRAO
staff\footnote{http://www.vla.nrao.edu/astro/calib/polar/}).  However,
the polarization position angle for this source can change by several
tens of degrees in two weeks, so the absolute polarization position
angles derived (the electric vector position angles; EVPAs) should be
treated with caution.  Nevertheless, the magnitude of polarization, $P
= \sqrt{Q^2+U^2}$ (where $Q$ and $U$ are the polarized flux densities
in the Stokes $Q$ and $U$ parameters) is reliable.

\begin{figure}
\epsfig{figure=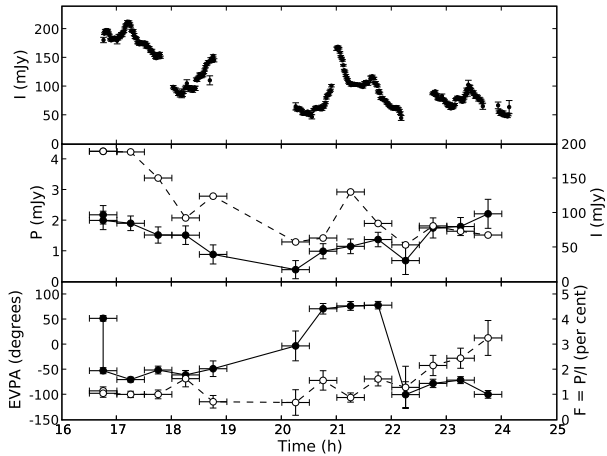,width=\columnwidth}
\caption{\label{fig:polarization}The polarization parameters of Cygnus
  X-3 at 43\,GHz.  Top panel shows the full total intensity 43-GHz VLA
  lightcurve, middle panel shows the total intensity lightcurve binned
  in half-hour-long intervals (open circles, dashed line, right hand axis)
  and the linear polarization lightcurve (filled circles, solid line,
  left hand axis) binned to the same resolution.  Bottom panel shows
  the electric vector position angle (filled circles, solid line, left
  hand axis) and fractional polarization in per cent (open circles,
  dashed line, right hand axis) as a function of time.  In the lower
  two panels, the two data points for the first time bin denote the
  two distinct peaks seen in the polarization image.}
\end{figure}

At 15\,GHz, no evidence for linear polarization at the source position
was seen, down to an r.m.s.\ level of 0.22\,mJy\perbeam\ for a single
20-min scan and 0.06\,mJy\perbeam\ for the full dataset.  For a mean
15-GHz flux density of 177.0\,mJy, this corresponds to a $5\sigma$
upper limit of 0.17 per cent on the linear polarization.  At 43\,GHz,
however, significant linear polarization was detected, at levels of up
to 2.5\,mJy.  The level of linear polarization initially decreases
in line with the total intensity, but then rises during and after the
flare at $\sim21$\,h (Fig.~\ref{fig:polarization}).  No circular
polarization was detected at either frequency to a $5\sigma$ limit of
0.6\,mJy\perbeam.

Linear polarization has occasionally been detected in previous giant
flares of Cyg X-3 \citep[e.g][]{Sea74,Led76}, up to levels of $\sim20$
per cent, with the fractional polarization increasing with frequency.
However, the majority of observations have shown no significant
polarization, with upper limits of a few per cent
\citep[e.g.][]{Bra72,Hje72}, so there are relatively few datasets
available for comparison.  \citet{Tud07} used the European VLBI
Network (EVN) to resolve polarized emission from Cyg X-3, finding a
maximum fractional polarization of 25 per cent at the boundary where
an ejected knot was interacting with its environment.  For the derived
rotation measure (RM) of $-1233$\,rad\,m$^{-2}$ \citep{Led76}, we
would expect a rotation of $-0.5$\,rad at 15\,GHz, and $-0.06$\,rad at
43\,GHz.  This RM was derived during giant outbursts, when the
emitting regions had propagated further from the core than those we
see during these small flaring events.  It is likely therefore, that
there is some extra Faraday rotation in the inner region, causing
depolarization at the lower frequency either due to averaging within
the synthesised beam of the observations or along the line of sight.
However, since the theoretical maximum fractional polarization for an
optically thick source is $\sim 12$ per cent, as compared to $\sim 70$
per cent for an optically thin source, we would in any case expect a
lower degree of polarization at 15\,GHz, where the optical depth
is greater.  The beam and line-of-sight depolarization then reduce
the observed degree of polarization still further, below our detection
threshold.

At 43\,GHz, the polarization position angle is stable to within
$\sim20$\degr\ except during the flare at $\sim21$\,h, which causes a
significant, albeit temporary, rotation of the polarization position
angle, while the degree of linear polarization increases slightly.  If
this flare is due to the formation of internal shocks in the flow (see
Section \ref{sec:modelling}), the resulting compression of the
magnetic field at the shock front could be responsible for the
rotation of the polarization position angle and the increase in
polarized intensity.  Bearing in mind the uncertainty in absolute
position angle calibration, we see the mean EVPA evolve from
$-58^{\circ}\pm10^{\circ}$ at the start of the observations, through
$76^{\circ}\pm4^{\circ}$ during the flare at 21\,h, back to
$-84^{\circ}\pm15^{\circ}$ at the end of the observations (as shown in
Fig.~\ref{fig:polarization}).  For optically-thin emission, the
measured EVPA should be perpendicular to the magnetic field
orientation.  Previous high-resolution observations
\citep{Mio01,Mil04,Tud07} have shown a north-south jet axis with a
position angle close to $0^{\circ}$.  Thus a field principally aligned
along the jet axis should have an EVPA of $\sim90^{\circ}$, while
compression due to a shock viewed sideways should give an EVPA of
$\sim0^{\circ}$.  While the EVPA at the end of the observations is
consistent with a field aligned along the jet axis, there are several
possible explanations for the intermediate position angles seen
earlier.  In the absence of precession, either there is additional
Faraday rotation beyond the $3^{\circ}$ expected from the previously
derived RM, the EVPA calibration is inaccurate, or we are seeing
emission from a superposition of emitting regions, some with a
longitudinal field configuration (upstream of the shock front), and
some from the compressed fields at the shock front.  Alternatively, if
we are seeing the shock front close to face on, as has been proposed
for Cygnus X-3 \citep{Lin07}, owing to the small angle between the jet
axis and the line of sight \citep{Mio01,Mil04}, we would not expect to
see an increase in the degree of ordering of the field.  In the
absence of data at different frequencies, or at higher spatial and
time resolution, it is not possible to differentiate between these
competing explanations.

The rise in fractional polarization towards the end of the observation
could be explained in part by the expansion of the emitting region
following the large flare at the start of the observations.  As the
source becomes more optically thin, the theoretical maximum fractional
polarization increases.  Furthermore, the density of the stellar wind
of the companion star decreases as the ejecta move outwards, reducing
the effect of Faraday depolarization in the wind.  Alternatively, a
change in the relative dominance of a depolarized core and
highly-polarized ejecta, as seen in XTE J\,1748-288 \citep{Bro07},
could be behind the increase in fractional polarization (consistent
with the hypothesis of multiple emitting regions proposed to explain
the observed EVPAs).  Multifrequency data or resolved VLBI
observations would be required to test these hypotheses, while higher
sensitivity would allow us to make higher-time resolution lightcurves
in linear polarization, and more accurately probe how well the
polarized emission followed the total intensity.  Upcoming facilities
such as EVLA and eMERLIN will provide the required increases in
sensitivity and fractional bandwidth to make such observations.

\subsection{Imaging}
\label{sec:25jan_imaging}
Imaging at 43\,GHz with the VLA in A-configuration gives an angular
resolution of order 50\,mas.  This would be sufficient to resolve
arcsecond-scale extensions such as those seen by \citet{Mar01}, if
present during our observations.  However, given the time-variable
nature of the source, the variability had to be removed prior to
imaging the source in order to search for any potential faint
extensions.  Imaging without removing the variability showed no
extension to an r.m.s.\ level of 1.8\,mJy\perbeam.  An automated
procedure was written using {\sc ParselTongue}, a Python interface to
{\sc aips}, to remove the time-variable core by making images of 1-min
time chunks, and subtracting the fitted core component in the {\it
uv}-plane before recombining and imaging the core-subtracted data from
all time intervals.  At 15\,GHz, there was no observable extension down
to an rms of 0.6\,mJy\perbeam.  The subtraction did not work as well
at 43\,GHz, possibly due to atmospheric distortions shifting the
source position slightly.  The central source was not fully removed,
although no evidence for extension was seen down to an rms of
0.2\,mJy\perbeam.

The different proper motion measurements quoted in the literature
\citep{Mio01,Mar01,Mil04} all predict that a flare would take
several days to become sufficiently extended to be resolved at 15\,GHz
by the VLA in its A-configuration.  The 400-mJy flare at the start of
our observations could not therefore have been resolved.  The Ryle
Telescope monitoring
data\footnote{http://www.mrao.cam.ac.uk/$\sim$guy} show no flares with
15-GHz flux densities exceeding 250\,mJy between the 2001 September
outburst \citep{Mil04} and the beginning of our observations.  The
lack of extended emission down to our rms limit of 0.2\,mJy then
constrains the e-folding time for the decay of any low-level flares to
be $<5.2$\,d, and that for the 2001 September giant flare to be
$<12.5$\,d.

\subsection{Spectra}
\label{sec:fittedspectra}
\begin{figure}
\epsfig{figure=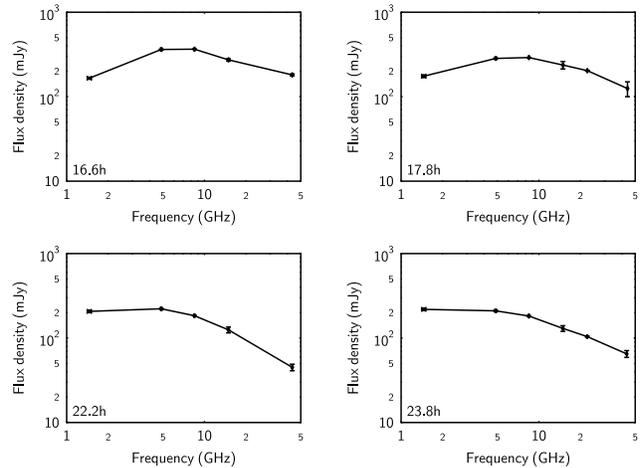,width=\columnwidth,angle=0,clip=}
\caption[Broadband radio spectra]{\label{fig:radio_spectra}Broadband
  radio spectra for the four epochs when the lightcurve was
  well-sampled.  The times (IAT hours relative to MJD\,52299.0) of the
  observations are noted in the bottom left of each plot.  Note the
  turnover moving to lower frequency and flux density with time.}
\end{figure}

The overall radio spectrum was sufficiently well-sampled on four
occasions over the course of the January 2002 observing run to
generate broadband spectra (Fig.~\ref{fig:radio_spectra}).
The spectra are consistent with canonical synchrotron spectra affected
by absorption at the lower frequencies.  The overall flux density
decreases with time and the spectral turnover moves to lower
frequencies.  This is the behaviour expected for the decay of the
400-mJy flare at the start of the observations, as the ejecta move
outwards from the core and expand.  The lower-level variability seen
in Fig.~\ref{fig:variability} is a secondary effect superposed on the
general decrease, indicative of multiple emitting components in the
source.

\section{Opacity effects}
\label{sec:opacity}
As noted in Section \ref{sec:lcs}, the lower-frequency lightcurve is a
smoothed and delayed version of the emission at the higher frequency, a
clear indication of the existence of opacity effects, as commonly
observed in flaring sources such as AGN \citep[e.g.][]{All85},
supernovae \citep[e.g.][]{Wei86}, and gamma-ray bursts
\citep[e.g.][]{Sod06}.  Since the optical depth for both free-free
absorption and synchrotron self-absorption decreases with increasing
frequency, we probe more compact regions at high frequencies, with
lower-frequency emission being delayed until the source has either
expanded or moved out from behind the absorbing medium and the optical
depth has decreased to order unity.

\citet{Mol84} observed similar low-level activity in Cyg X-3 with
small flares which peaked later and with smaller amplitudes at lower
frequencies.  They measured a delay of $15\pm5$\,min between 22 and
15\,GHz, increasing to $239\pm7$\,min between 15 and 1.4\,GHz.  They
also claimed evidence for a periodicity in the range 4.8--5.1\,h,
although such a periodicity has not since been detected
\citep[e.g.][who note that in fact the amplitude of any periodic
phenomenon in the data of \citeauthor{Mol84} is less than
30\,mJy]{Joh86}.  To search for any periodicity in our data, we
constructed the power spectra of the lightcurves at the two
frequencies, which are shown in Fig.~\ref{fig:powerspectra}.

\subsection{Power spectra}
\label{sec:powerspectrum}
\begin{figure}
\epsfig{figure=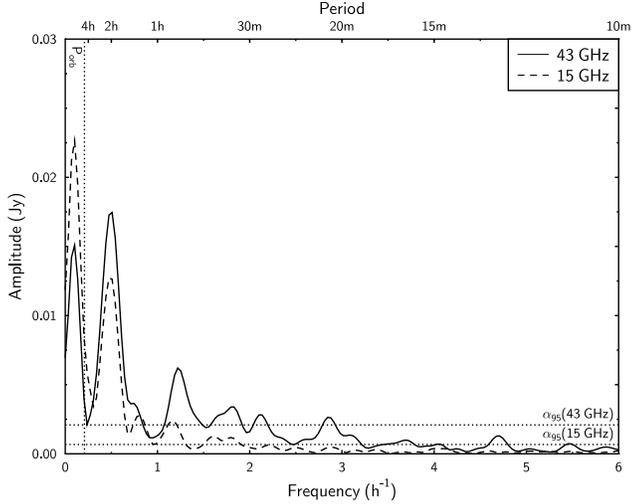,width=\columnwidth,clip=}
\caption[Power spectra of 15- and 43-GHz
  data]{\label{fig:powerspectra}Power spectra of 15- (dashed) and
  43-GHz (solid) data.  The vertical dotted line shows the 4.8-hour
  orbital period of the system.  The labelled horizontal dotted lines
  show the 95 per cent significance levels, $\alpha_{95}$, for the 15-
  and 43-GHz data.}
\end{figure}

Since the data were not evenly sampled in time, in order to construct
the true power spectra of the lightcurves, we implemented a
one-dimensional version of the CLEAN algorithm, as described by
\citet{Rob87} and used in standard data reduction procedures for
aperture synthesis radio telescopes.  We ascertained the significance
level of the peaks in the CLEAN spectra by performing Monte Carlo
simulations, randomly stripping out half the data points in the time
series and calculating the CLEAN spectra using the remaining data
\citep{Hes02}.  The 95 per cent confidence limit $\alpha_{95}$ was set
at the point below which lay 95 per cent of the sorted, concatenated
CLEAN spectra from all 1000 Monte Carlo iterations.  The final CLEAN
spectra of the 43 and 15\,GHz data are shown in
Fig.\,\ref{fig:powerspectra}, together with their respective
significance levels $\alpha_{95}$ of 2.1\,mJy and 0.67\,mJy
respectively.

There is no significant power on periods shorter than about 20\,min at
43\,GHz, and about 30\,min at 15\,GHz.  These minimum variability
timescales provide approximate constraints on the size scales for the
$\tau=1$ surface of the emitting region at the two frequencies, equal
to 2.4 and 3.6\,AU respectively (assuming material to be travelling at
$c$).  At 10\,kpc \citep{Dic83}, this corresponds to a size of
0.2--0.3\,mas.  This may be compared with the predicted scattering
size of $448(\nu/1{\rm GHz})^{-2.09}$\,mas \citep{Mio01}, which
equates to 0.2\,mas at 43\,GHz and 1.6\,mas at 15\,GHz.  At
15\,GHz, we are probing regions smaller than can be accessed with
VLBI imaging observations.

The power at 15\,GHz is uniformly lower than at 43\,GHz, owing to the
smoothing out of the lower-frequency emission due to the larger size
of the emitting region.  The most significant peaks in the power
spectra correspond to a period of about 2\,h at both frequencies.
The lowest-frequency peaks correspond to the overall decrease in the
mean flux density with time, and are not sampled sufficiently well to
correspond to a believable periodicity.  There is no significant peak
at the orbital period (4.8\,h) at either frequency, consistent with
the findings of \citet{Ogl01b}.

\subsection{Delay between the frequencies}

To quantify the time delay between the two frequencies, we
cross-correlated the lightcurves using the `locally-normalised
discrete correlation function' (LNDCF) of \citet{Leh92}.  This
accounts for the uneven sampling of the data and the non-stationary
nature of the time series, preventing the correlation function from
being dominated by interpolations (as is the case for the standard
correlation function method) and avoiding the introduction of sampling
artifacts \citep{Mol86,Pet93}.

\begin{figure}
\epsfig{figure=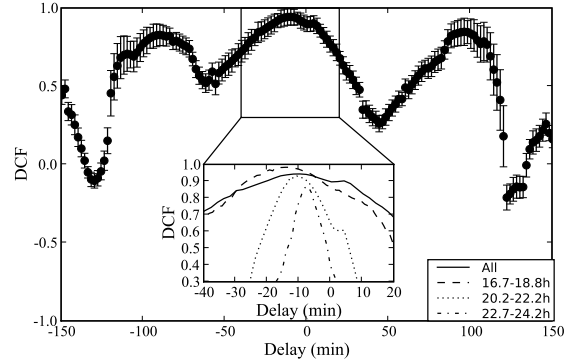,width=\columnwidth}
\caption{\label{fig:ccf}Cross-correlation between the 15-GHz and the
  43-GHz lightcurves.  Negative delays indicate that the 15-GHz
  lightcurve lags the 43-GHz lightcurve.  Lag bins are of length
  2\,min.  Inset shows the peak of the cross-correlation function for
  the three different sections of data, plotted without error bars for
  clarity.  The solid line is for the entire data set, the dashed line
  is for Section 1, the dotted line is for Section 2 and the
  dot-dashed line is for Section 3.}
\end{figure}

The derived cross-correlation function is shown in Fig.~\ref{fig:ccf}.
It peaks when the 15-GHz emission lags that at 43\,GHz by
$9.5\pm1.0$\,min, broadly consistent with the $15\pm5$\,min delay
between 22 and 15\,GHz found by \citet{Mol84}.  Splitting the data
into three sections (16.7--18.8\,h, 20.2--22.2\,h and 22.7--24.2\,h
UT, as labelled in Fig.\,\ref{fig:variability}), the lag was found to
decrease with time, from $13.2\pm1.0$\,min in the first section to
$7.0\pm1.0$\,min for the final section.  The peak in the
cross-correlation function also becomes narrower, with a diminishing
amplitude with time, as shown in Fig.\,\ref{fig:ccf}.  This is
suggestive of decreasing opacity through the flaring sequence, a trend
also seen during the multi-flare outburst sequence of 1994
February--March \citep{Fen97}, and explained as a decreasing amount of
entrained thermal material with time over the course of the outburst.

In an attempt to better understand the relation between the two
lightcurves, we now attempt to fit the data with a shock-in-jet model.

\section{The shock-in-jet model}
\label{sec:modelling}

Shock-in-jet models have been successfully fitted to the lightcurves
of the 1994 outburst of GRO J\,1655-40 \citep{Ste03} and to a series
of small-scale oscillation events in GRS\,1915+105 \citep{Tur04}.  It
was suggested that the same model could be used to fit the giant
outbursts of the latter source, by simply scaling the timescales, peak
flux densities and frequencies of the individual events.
\citet{Lin07} applied the shock-in-jet model to Cygnus X-3, finding a
good fit to the data from both the 2001 giant outburst and a sequence
of intermediate-level flares from 1994.  To ascertain whether such a
shock model was appropriate to the smaller-scale events we observed in
2002, and to derive some of the physical parameters of the outburst
for comparison with the stronger flares seen in this system, we fitted
the radio lightcurves with a shock-in-jet model.

\subsection{Outline of the model}
An analytic model for the evolution of a shock wave propagating down
an adiabatically-expanding conical jet was originally derived by
\citet{Mar85}.  This model was generalised by \citet{Tur00}, and
further refined for application to microquasar systems by
\citet{Tur04} and \citet{Lin07}.  According to this model, a shock
wave propagating down a jet evolves through three different phases, in
which the dominant energy loss mechanism evolves from inverse Compton
losses to synchrotron emission, and finally adiabatic expansion.  In
the inverse Compton stage, the peak emission frequency decreases while
the peak flux density rises.  Once the photon energy density becomes
equal to the magnetic energy density, synchrotron losses take over,
and the spectral peak moves to lower frequency at roughly constant
flux density.  Finally, during the adiabatic phase, both the turnover
frequency and flux density decrease with time.

To fit the observed lightcurves with this model, each peak in the
lightcurve is associated with the evolution of a single shock front.
The fitted start time of an event, $T_0$, is the time of onset of the
shock, which has been found to correspond fairly closely to the
zero-separation time of individual VLBI components in AGN jets
\citep{Tur99,Lin06}.  All shocks are assumed to be self-similar, and
the characteristics of an individual event (hereafter referred to as
the specificities) are set only by a shift in peak flux density,
frequency and duration with respect to the average outburst, to
minimize the number of free parameters of the model.

A detailed description of the model is given by \citet{Tur00},
although we have incorporated a number of refinements in fitting the
lightcurves presented in Section \ref{sec:obs}.  \citet{Bjo00}
modified the shock model to make the rise in flux density in the
initial Compton stage less abrupt than was predicted by the original
model of \citet{Mar85}.  With this modification, it was found to be
difficult to differentiate the Compton and synchrotron stages
\citep{Lin07}, particularly in the absence of high-frequency infrared
data, so we omitted the synchrotron stage from our modelling.  We
assumed a very simple adiabatically-expanding, conical and
non-accelerating (constant Doppler-factor) jet.  We also assumed a
homogeneous synchrotron source, and that all the emission originates
from the approaching jet, since \citet{Lin07} found that accounting
for the receding jet in this source did not improve the fit to the
major and intermediate outbursts of Cygnus X-3, most likely due to the
small angle of the flow to the line of sight.

This model differs from a standard plasmon model \citep{van66}
principally by accounting for the rise phase of the flare in which
radiative (Compton or synchrotron) losses dominate over adiabatic
losses.  In this initial radiative phase, the electrons cannot move
far from the shock front before losing energy.  The thickness of the
emission region behind the shock front then varies close to
quadratically with the jet radius $R$, with the exact scaling
depending on the parameters of the model \citep{Bjo00}.  Thus the
volume decreases more rapidly on tracing the emission back to the apex
of the jet, and due to the decreased source volume, the flux density
at optically thin frequencies (in the infrared) does not become
excessively high at small radii.  While we have no simultaneous
infrared data from 2002 January 25, and cannot mandate the use of a
shock model as opposed to a plasmon model, application of this
shock-in-jet model to previous outbursts of Cygnus X-3 \citep{Lin07}
and to GRS\,1915+105 \citep{Tur04} has shown that it can reproduce
both the infrared and radio lightcurves.  The second main difference
is that since the shock-in-jet model involves an underlying flow
expanding in two dimensions, the electron energy decreases more slowly
with time, as $E\propto R^{-2/3}$ rather than the $R^{-1}$ scaling
expected for a three-dimensional expansion of a discrete plasmon with
no turbulence.  This in turn leads to a slightly less steep decay of
the synchrotron self-absorption turnover flux density with frequency
and time.

\subsection{Fitting the lightcurves}
\begin{figure*}
\epsfig{figure=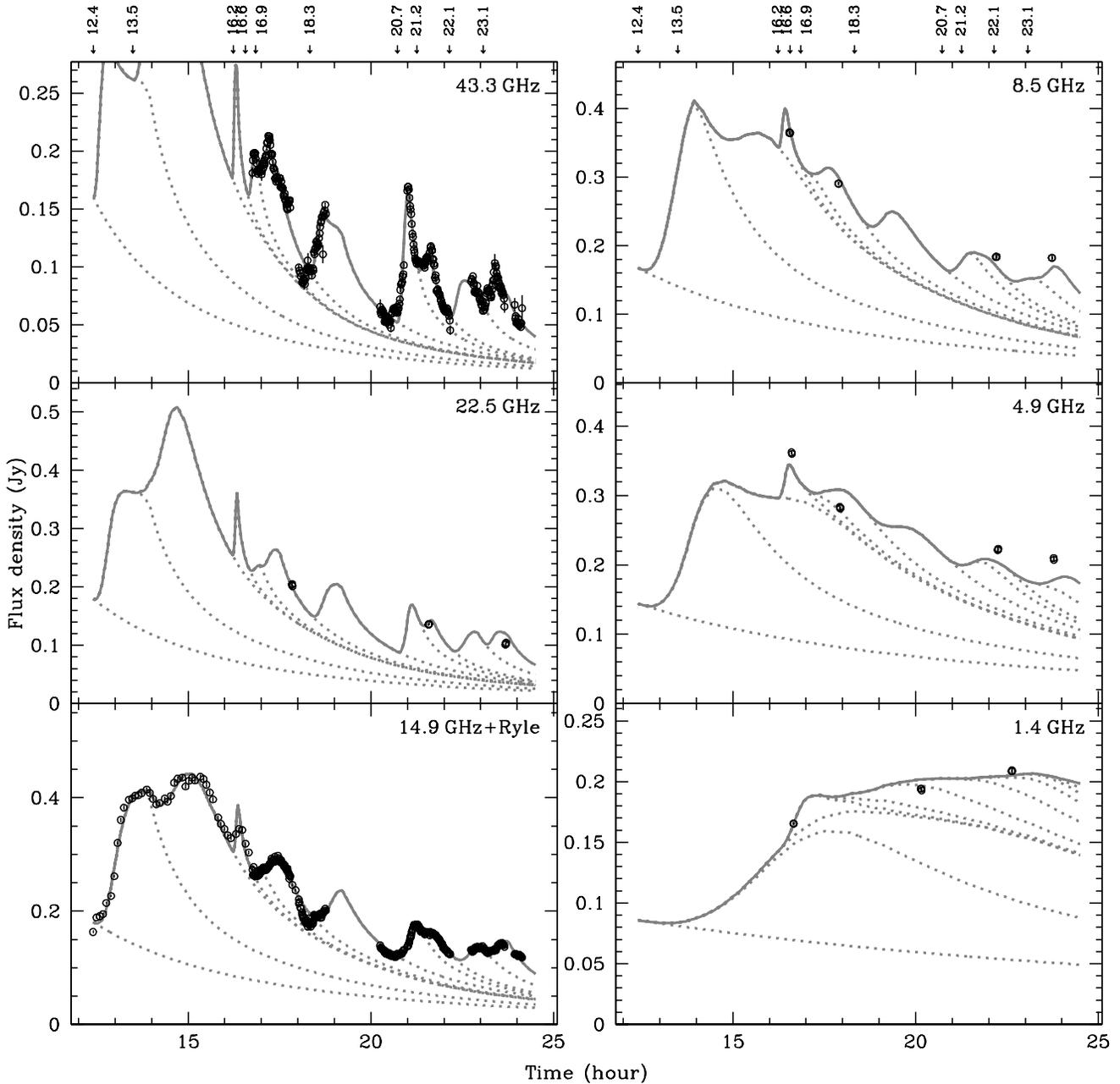,width=\textwidth}
\caption{\label{fig:shock_lc_fit}Best-fitting model lightcurves at
each frequency, with the data overplotted as open circles.  The 15-GHz
Ryle Telescope flux densities have been multiplied by a factor 1.25 to
bring them into agreement with the 14.9-GHz VLA data.  The different
outburst events are shown by dotted lines.  The frequency of each
lightcurve is marked in the top right of each panel.  The IAT start
times of the different outbursts are indicated by arrows at the top of
the figure.}
\end{figure*}

The radio lightcurves were decomposed into a series of 10 individual
events (each representing the development and propagation of a shock
within the flow), plus an initial decaying outburst representing
residual emission from events prior to the start of the observing
period.  All frequencies were fit simultaneously to determine both the
profile of the average outburst and the logarithmic shifts in flux
density ($\Delta\log S$), frequency ($\Delta\log\nu$) and time
($\Delta\log t$) of the transition from the Compton to the adiabatic
stage, which determine the specificities of the individual events.  It
was found that in the absence of infrared data to constrain the
high-frequency emission, there was significant degeneracy along the
adiabatic evolution track.  To discriminate between different
weakly-constrained solutions for the specificities of individual
outbursts, an extra constraint was placed on the fit parameters,
minimizing the combined scatter in the specificities $\Delta\log S$,
$\Delta\log\nu$ and $\Delta\log t$.  The model used 51 free parameters
(seven to specify the shape and evolution of the average event
profile, four to represent residual decaying emission at the start of
the observations, plus four per outburst to determine the
characteristics of each of the ten individual events), fitted to 700
data points.  The best fitting model, giving a reduced $\chi^2$ of
20.5 for the entire dataset, is shown for the different observing
frequencies in Fig.~\ref{fig:shock_lc_fit}.  The converged model
parameters are given in Table \ref{tab:model}, and the derived start
times and logarithmic shifts of the different events are listed in
Table \ref{tab:shifts}.

\begin{table}
\begin{center}
\begin{tabular}{lc}
\hline\hline
Parameter & Value \\
\hline
$s$ & 2.11 \\
$b$ & 1.19 \\
$t_{\rm p}$ (h) & 0.47 \\
$\nu_{\rm p}$ (GHz) & 26.9 \\
$S_{\rm p}$ (Jy) & 0.14 \\
$\nu_{\rm b}$ (THz) & $2.30$ \\
$t_{\rm f}$ (h) & 0.11 \\
\hline\hline
\end{tabular}
\end{center}
{\caption{\label{tab:model}The seven model parameters specifying the
    shape of the average outburst.  $s$ is the index of the electron
    energy spectrum, $b$ is the scaling of the magnetic field with
    distance $z$ along the jet ($B\propto z^{-b}$), $S_{\rm p}$ is the
    peak flux density at the transition to the adiabatic stage,
    $\nu_{\rm p}$ is the frequency at which the flux density peaks,
    and $t_{\rm p}$ is the time of the peak relative to the onset of
    the shock at time $T_0$.  $\nu_{\rm b}$ is the frequency of the
    high-frequency break in the spectrum where it steepens from
    $\nu^{-(s-1)/2}$ to $\nu^{-s/2}$, defined at the time of the peak,
    and $t_{\rm f}$ is the time at which the spectrum begins to
    flatten to $\nu^{-(s-1)/2}$ prior to the start of the adiabatic
    stage.}}
\end{table}

While inspection of Fig.~\ref{fig:shock_lc_fit} shows that the
model appears to reproduce the overall shape of the lightcurves fairly
well at 15 and 43\,GHz, the reduced $\chi^2$ for the overall fit is
relatively high.  It is possible that the error bars on the data
points may have been slightly underestimated.  The uncertainties on
the data are mainly statistical, although some extra systematic errors
of 1 per cent were added to the Ryle Telescope data and the
lower-frequency VLA data.  However, they do not account for the 3--5
per cent uncertainty in the absolute flux scale at the VLA, since an
overall gain error should affect all data points at a given frequency
and have no effect on the relative shape of the lightcurves, whereas
the discrepancies between the model and the data are in the fine
details of the lightcurves rather than their overall normalisations.
It is more likely that at these high frequencies, atmospheric
variations on a timescale shorter than the switching time between
calibrator and target caused extra scatter in the observed flux
density.

However, the high reduced $\chi^2$ value also reflects the fact that
the model is a fairly simplistic idealisation, which does not attempt
to reproduce in detail the exact shapes of individual events, and so
is unlikely to be able to account for all features of the observed
lightcurves.  We assumed a conical jet, a strong shock, no reverse
shock, and did not account for counter-jet emission or jet precession.
The slope of the electron spectrum $s$ was assumed to be constant with
time, and the ratio, $\nu_{\rm p}/\nu_{\rm b}$, of the peak frequency
to the high-frequency break of the optically-thin synchrotron spectrum
was assumed to be the same for all outbursts (albeit evolving with
time within an individual event).  Real magnetohydrodynamical jets are
likely to be far more complicated, with phenomena such as transverse
shocks, stationary knots, and Kelvin-Helmholtz instabilities, which we
cannot begin to account for in our self-similar analytic model.  The
fit is clearly sufficient to validate the zero-order picture of
adiabatically-expanding synchrotron plasma, although in the absence of
high-frequency millimetre or infrared data, discriminating between the
rise phases of a shock-in-jet model and a plasmon model is more
difficult, and deriving the details of any shock structure is clearly
impossible without detailed high-resolution VLBI imaging.

While the electron energy decays more slowly with radius in the
shock-in-jet model, the addition of turbulence into the plasma in a
plasmon model can slow the decay of the flux density as the plasmon
expands.  We can examine the expected synchrotron flux density during
the adiabatic decay phase for the two different models
\citep[e.g.][]{Lon94},
\begin{equation}
S_{\nu} \propto R^3\kappa B^{(s+1)/2}\nu^{-(s-1)/2},
\end{equation}
where $\kappa$ is the normalisation of the electron spectrum,
$N(E)\propto\kappa E^{-s}$, and $B$ is the magnetic field strength.
For the shock-in-jet model, $B\propto R^{-b}$ and $\kappa\propto
R^{-2(s+2)/3}$.  Thus, for the shock-in-jet model,
\begin{equation}
S_{\nu}\propto R^{[10-4s-3b(s+1)]/6}\nu^{-(s-1)/2}.
\end{equation}
From the values in Table 1, this scales as $S_{\nu}\propto R^{-1.59}$.
For the plasmon model, $B\propto R^{-2+\eta}$, where $\eta$ is a
factor describing the degree of turbulence in the plasma, typically of
order 1, with $\eta=0$ corresponding to no strong turbulence
\citep{Lan63}.  The first adiabatic invariant, $p_{\perp}^2/B$, is
conserved during the expansion, where $p_{\perp}$ is the component of
the particle momentum, $p=E/c$, perpendicular to the magnetic field.
If the particle distribution is isotropic throughout the expansion,
then $p_{\perp}^2 = 2p^2/3$, such that $E\propto R^{(-2+\eta)/2}$.
Assuming the number of particles in a plasmon, $VN(E)\,\dd E = V\kappa
E^{-s}\,\dd E$, is conserved, we then find that $\kappa\propto
R^{-[2+s+\eta(1-s)/2]}$, such that
\begin{equation}
S_{\nu}\propto R^{s(\eta-2)}\nu^{-(s-1)/2}.
\end{equation}
For the same value of $s$ as fitted for the shock-in-jet model, a
value $\eta=1.24$ would then provide the same scaling during the decay
phase.  Thus we cannot use the decay phases of the outbursts to
conclusively discriminate between the two classes of model.
Higher-time resolution polarization data would help to discriminate
between the two types of flow, since the compression and consequent
ordering of the magnetic field at the shock front at the onset of an
event would lead to a higher degree of polarized emission and a
rotation of the mean position angle of the field, from one aligned
parallel to the jet to one aligned perpendicular to it \citep{Hug89a}.
While there are hints of a rotation of the EVPA during the flare at
21\,h (Fig.~\ref{fig:polarization}), the fractional polarization does
not increase significantly, and higher time resolution (i.e.\ higher
sensitivity) would be required to verify this signature of a
propagating shock front.

Nevertheless, owing to the good overall match to the shape of the
lightcurves, we will assume that the basic zero-order properties of
this model are correct, and go on to use the fitted model parameters
to constrain the differences between the 2002 events and the
previously-studied outburst sequences of 1994 and 2001.

\begin{figure}
\epsfig{figure=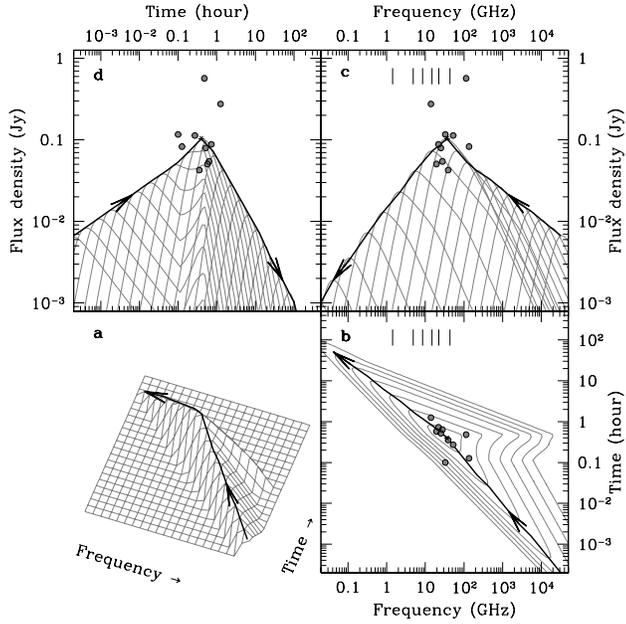,width=\columnwidth}
\caption{\label{fig:shock_evolution}The evolution of the average
  outburst in frequency-time-flux density space. (a) indicates the
  path followed by the self-absorption turnover, and (b), (c) and (d)
  show the projections onto the frequency-time, frequency-flux density
  and time-flux density axes respectively.  The contours in (b) are
  points of equal flux density, spaced by 0.3\,dex.  In (c), the grey
  lines show individual spectra at times spaced by 0.3\,dex.  In (d),
  they show lightcurves spaced by 0.3\,dex in frequency.  The thick
  black lines show the path followed by the self-absorption turnover,
  and the grey dots show the time, flux density and frequency of the
  peak emission of the individual outbursts, each of which is
  represented by a self-similar scaling of the average outburst.  The
  vertical black lines at the top of (b) and (c) show the observing
  frequencies.  The two outlying points in (c) and (d) are the two
  earliest flares (starting at 12.38 and 13.47\,h respectively), which
  are constrained only by the Ryle Telescope data and the later
  low-frequency points.}
\end{figure}

The shape of the fitted average outburst in the three-dimensional
parameter space defined by peak flux density, frequency and time after
onset is shown in Fig.~\ref{fig:shock_evolution}.  Also shown are the
locations of the peaks of the individual outbursts, calculated from
the average outburst values and the logarithmic shifts in time,
frequency and flux density of the individual events (given in Table
\ref{tab:shifts}).  While there is significant scatter in flux
density, they are well-aligned in the frequency-time plane along the
locus of the spectral turnover (Fig.~\ref{fig:shock_evolution}),
suggesting that the turnover always follows a similar track in
frequency and time.  The scatter in flux density suggests that another
factor is at work in determining the brightness of a given outburst,
possibly the normalisation of the electron spectrum, $\kappa$.  Such
an anticorrelation between frequency and timescale has been previously
noted in 3C\,273 (T\"urler, Courvoisier \& Paltani 1999) and Cygnus
X-3 \citep{Lin07}, but not in 3C\,279 \citep{Lin06}.

\begin{table}
\begin{center}
\begin{tabular}{lcccc}
\hline\hline
Epoch & $T_0$ & $\Delta\log S$  & $\Delta\log\nu$  & $\Delta\log t$\\
\hline
1994 & 2.95\,d & 1.53 & -0.59 &  1.58 \\
& 5.80\,d & 0.92 & -0.65 & 1.62 \\
& 7.25\,d & 0.73 & -0.16 & 1.55 \\
& 9.06\,d & 0.53 & -0.29 & 2.18 \\
& 11.44\,d & 0.66 & -0.94 & 1.18 \\
& 12.31\,d & 1.35 & -0.62 & 1.13 \\
& 14.16\,d & 0.92 & -1.28 & 1.75 \\
& 16.19\,d & 1.33 & -0.72 & 1.06 \\
& 18.53\,d & 0.92 & -0.94 & 1.94 \\
& 20.37\,d & 1.65 & -0.87 & 1.14 \\
& 22.89\,d & 0.90 & -1.15 & 1.63 \\
& 23.72\,d & 1.20 & -1.28 & 1.97 \\
& 30.09\,d & 0.50 & -0.66 & 1.43 \\
\hline
2001 & 2.64\,d & 1.98 & -0.76 & 1.76 \\
& 4.51\,d & 1.80 & -1.14 & 1.94 \\
& 4.64\,d & 1.70 & -0.73 & 1.78 \\
\hline
2002 & 12.38\,h &  0.43 & -0.42 &  0.50 \\    
& 13.47\,h &  0.74 &  0.49 &  0.08 \\
& 16.21\,h &  0.05 & -0.04 & -0.60 \\
& 16.59\,h & -0.10 &  0.57 & -0.50 \\
& 16.88\,h & -0.11 & -0.16 &  0.11 \\
& 18.31\,h & -0.07 & -0.22 &  0.26 \\
& 20.71\,h & 0.04 & 0.16 & -0.17 \\
& 21.24\,h & -0.39 & 0.03 & -0.05 \\ 
& 22.13\,h & -0.28 & -0.12 &  0.20 \\
& 23.06\,h & -0.31 & -0.28 &  0.16 \\
\hline
\end{tabular}
\end{center}
{\caption{\label{tab:shifts}The fitted start times and logarithmic
    shifts in flux density, frequency and time, for the three outbursts of Cyg
    X-3, quoted relative to the average outburst derived for the 2002
    dataset.}}
\end{table}

\subsection{Comparison to previous outbursts}
A similar shock model was used by \citet{Lin07} to fit the lightcurves
of the 1994 February-March and 2001 September outburst sequences of
Cygnus X-3.  Both flaring sequences were significantly brighter than
the events observed in this paper.  In order to make valid comparisons
between the different outbursts, we re-fitted those datasets using the
average outburst and model parameters found from the 2002 January 25
dataset.  This provided consistent specificities for the individual
outbursts of the different flaring sequences.  The new logarithmic
shifts, which gave reduced-$\chi^2$ values for the 1994 and 2001
datasets of 11.8 and 3.0 respectively \citep[as compared to values of
10.2 and 1.8 from the original fits by][]{Lin07}, are given in Table
\ref{tab:shifts}.  We note that the sparse infrared data from the 1994
outburst sequence were not well fitted by these model parameters, since
there were no infrared data available in 2002 to constrain the
high-frequency spectral evolution.  However, since our goal was to
compare the radio emission from different outbursts in a consistent
fashion, we used the average outburst derived from the 2002 data to
fit all three flaring sequences, owing to the simplicity of the model
\citep[an adiabatically-expanding conical jet, following the original
assumptions of][]{Mar85}.

\begin{figure}
\epsfig{figure=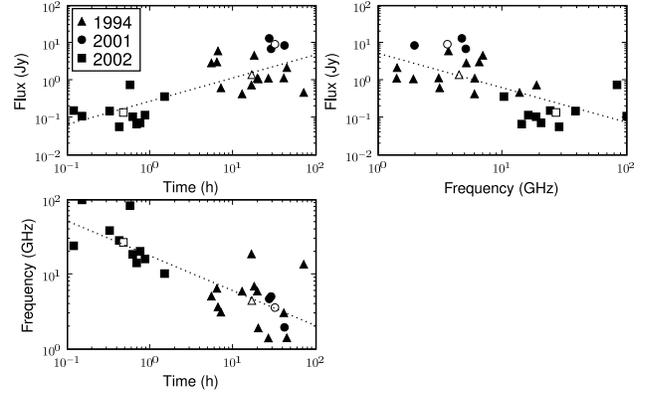,width=\columnwidth}
\caption{\label{fig:shock_parms}Peak flux density, frequency and time
  after onset of the shock, for each of the individual outbursts of
  Cygnus X-3 in 1994 (filled triangles), 2001 (filled circles) and
  2002 (filled squares), plotted against one another.  The open
  markers denote the position of the average outburst for each
  dataset.  The dotted lines show the least-squares fits for the
  scaling of the different parameters with one another; $S_{\rm
  p}\propto t_{\rm p}^{0.62}$, $S_{\rm p}\propto\nu_{\rm p}^{-0.93}$,
  and $\nu_{\rm p}\propto t_{\rm p}^{-0.47}$.}
\end{figure}

In Fig.~\ref{fig:shock_parms}, we use the specificities of the
individual outbursts (Table \ref{tab:shifts}) to plot the times,
frequencies and flux densities of the peak emission against one
another for all three datasets.  The peak frequency is anticorrelated
with the outburst timescale, not just for the 2002 data, but also
between the different outburst sequences.  Events that evolve faster
peak at higher frequencies (the calculated Spearman rank correlation
coefficient is $-0.79$, chance probability $<2\times10^{-6}$), as
found by \citet{Lin07}.  The timescale of the outburst also shows a
positive correlation with the peak flux density (correlation
coefficient 0.69, chance probability $<10^{-4}$), and there is an
anticorrelation between the peak frequency and peak flux density
(correlation coefficient $-0.68$, chance probability
$<2\times10^{-4}$).  Thus the shorter, higher-frequency-peaking
outbursts also tend to be fainter.  Fitted scalings of the three
parameters $S_{\rm p}$, $\nu_{\rm p}$ and $t_{\rm p}$ with one another
are given in Table \ref{tab:scalings}.

The frequency-timescale correlation could potentially be explained by
travel time arguments.  If the time of peak emission is related by
travel time to the distance from the jet base, then events occurring
closer to the core peak earlier and, since they have expanded less, do
so at higher frequencies.  Self-similar expansion implies that these
events evolve fastest.

\subsection{Physical parameters}
Assuming that the time of peak flux density can be approximated by
travel time down the jet, this occurs at a distance $z_{\rm p}
=\gamma\beta\delta ct_{\rm p}$ downstream, at which point the jet
radius is $R_{\rm p}=z_{\rm p}\tan\phi$ (assuming a cone opening angle
of $\phi=2$\degr\ in all cases).  The source angular size is given by
$\theta_{\rm src} = 2R_{\rm p}/d$, where $d$ is the source distance
\citep[taken to be 10\,kpc for Cyg X-3;][]{Dic83}.  Given the angular
size $\theta_{\rm src}$, turnover frequency, $\nu_{\rm p}$, and flux
density, $S_{\rm p}$, and the electron index $s$, with the assumption
of a homogeneous spherical source with a power-law electron energy
spectrum, then the magnetic field $B$, and the normalisation of the
electron spectrum, $\kappa$, can be calculated at the turnover using
the equations of \citet{Mar87}, as previously done for a sample of
Galactic and AGN jets by \citet{Tur07}.  While for a shock-in-jet
model the source has a cylindrical slab-like geometry rather than
being spherical, the depth scales with $R$ during the self-similar
adiabatic phase, so approximating the source as a sphere will be in
error by the ratio of the slab thickness $x$ to the jet radius $R$, a
factor $\xi\sim0.005h^{-5}$ \citep[as derived for the case of 3C\,273;
][]{Mar85}, where $h=H_0/(100{\rm \,km\,s}^{-1}{\rm \,Mpc}^{-1})$ and
$H_0$ is the Hubble constant.  For the currently-accepted value of
$h=0.732$ \citep{Spe07}, $\xi\sim0.024$.  \citet{Mar85} also
approximated the density of the shocked region as being constant up to
a distance $x=\xi R$ behind the shock front, at which point it would
abruptly drop to zero.  They found that this should not introduce
significant errors into the proportionalities for their equations.
The assumption of homogeneity of the emitting region was further
justified by \citet{Bjo00}, and should certainly be sufficient to
derive the scalings between physical parameters which we now present.
Adapting the equations of \citet{Mar87} to account for the
non-spherical geometry, and assuming we see the emitting region with
the shock face-on \citep{Lin07}, the magnetic field is given by
\begin{equation}
B_{\rm p}=10^{-5}b(\alpha)\theta_{\rm src}^4 \nu_{\rm p}^5 S_{\rm p}^{-2}
\delta \quad{\rm G},
\label{eq:magfield}
\end{equation}
the electron normalisation is
\begin{multline}
\kappa_{\rm p} = 10^6 n(\alpha)d^{-1}\xi^{-1}\theta_{\rm src}^{-(4\alpha+7)} \nu_{\rm
  p}^{-(4\alpha+5)} S_{\rm p}^{2\alpha+3} \\
\times \delta^{-2(\alpha+2)}  {\rm erg}^{2\alpha}{\rm cm}^{-3},
\label{eq:enorm}
\end{multline}
and the optical depth to synchrotron self-absorption at the turnover
frequency is
\begin{equation}
\tau_{\rm p}(\alpha) = c_2(\alpha)\kappa_{\rm p} B_{\rm p}^{(3+2\alpha)/2} \nu_{\rm
  p}^{-(5+2\alpha)/2} \xi \theta_{\rm src} d\delta^{(5+2\alpha)/2},
\label{eq:selfabs}
\end{equation}
where the source size $\theta_{\rm src}$ is given in mas, the peak
emission frequency $\nu_{\rm p}$ in GHz, the peak flux density $S_{\rm
p}$ in Jy and the source distance $d$ in kpc.  $\alpha=(s-1)/2$ is the
spectral index, and $n(\alpha)$ and $c_2(\alpha)$ are functions
depending logarithmically on $\alpha$, with values of 0.27 and
$1.2\times10^{17}$ respectively for $\alpha=0.5$.  $b(\alpha)$ is a
slowly-varying function of $\alpha$ with a value of 3.2 for
$\alpha=0.5$.  Away from the peak, the optical depth scales with
frequency as $\tau \propto \nu^{-(2\alpha+5)/2}$.

\begin{figure}
\epsfig{figure=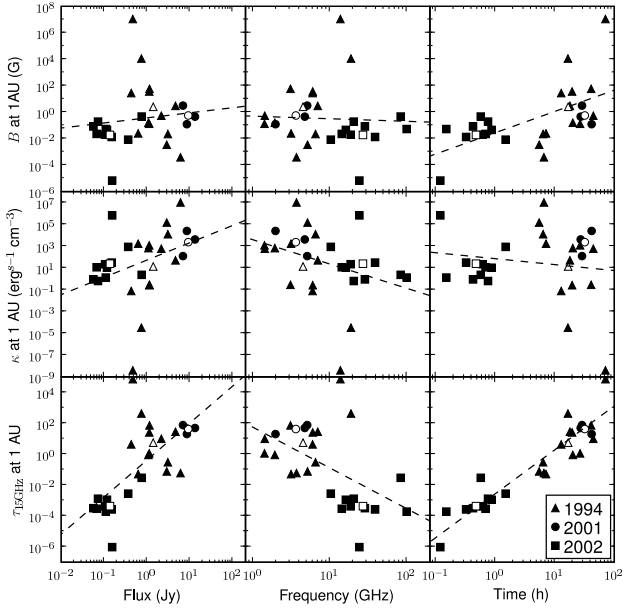,width=\columnwidth}
\caption{\label{fig:physparms}Magnetic field (upper panels), electron
  normalisation (middle panels), and 15-GHz optical depth to
  synchrotron self-absorption (lower panels) scaled to a jet radius of
  1\,AU, and plotted against peak flux density, frequency and time for
  each of the individual outbursts of Cygnus X-3 in 1994 (filled
  triangles), 2001 (filled circles) and 2002 (filled squares),
  together with the average outburst properties for each of the three
  flaring sequences (open symbols).  The dashed lines show the
  predictions for the scalings of the different parameters with flux,
  frequency, and time, according to the fitted coefficients shown in
  Table \ref{tab:scalings}.}
\end{figure}

To meaningfully compare the physical parameters for different
outbursts, they should be scaled to the same position $z$ along the
jet (i.e.\ to a fixed jet radius $R$).  The magnetic field and
electron normalisation scale as power laws with jet radius, $B\propto
R^{-b}$ and $\kappa\propto R^{-k}$.  $b$ is given in Table
\ref{tab:model}, and $k=2(s+2)/3$ for an adiabatic jet flow as assumed
here.  From the peak fluxes, frequencies and times of the individual
outbursts, the magnetic field, electron normalisation, and optical
depth to synchrotron self-absorption at a frequency of 15\,GHz were
calculated, scaled to the same jet radius of 1\,AU using the derived
values of $b$ and $k$, and plotted against peak flux density, $S_{\rm p}$,
frequency, $\nu_{\rm p}$, and time, $t_{\rm p}$ in
Fig.~\ref{fig:physparms}.  In all cases we assumed a distance of
10\,kpc \citep{Dic83}, a jet speed $\beta=0.63$, an inclination angle
of 10.5\degr\ to the line of sight and a jet half-opening angle of
2\degr \citep{Mil04}.  While we note that slightly different
parameters were found by \citet{Mio01}, we use a single set of
parameters here for consistency between outbursts.

\begin{table}
\begin{center}
\begin{tabular}{lcccccc}
\hline\hline
 & $S_{\rm p}$ & $\nu_{\rm p}$  & $t_{\rm p}$ & $B(1{\rm AU})$ &
 $\kappa(1{\rm AU})$ & $\tau(1{\rm AU})$\\
\hline
$S_{\rm p}$ & & -0.46 & 0.91 & [0.38] & 1.58 & 2.40\\
$\nu_{\rm p}$ & -0.93 & & -1.36 & [-0.19] & -2.23 & -2.64\\
$t_{\rm p}$ & 0.62 & -0.47 & & 1.61 & [-0.55] & 2.85\\
\hline
\end{tabular}
\end{center}
{\caption{\label{tab:scalings}Fitted scalings of the peak turnover
    flux density $S_{\rm p}$, frequency $\nu_{\rm p}$ and time $t_{\rm
    p}$ with one another and the resulting predictions, via equations
    \ref{eq:magfield}, \ref{eq:enorm}, and \ref{eq:selfabs}, for the
    scaling of the magnetic field, electron normalisation, and optical
    depth at 15\,GHz to synchrotron self-absorption, normalized to
    1\,AU, with the three parameters $S_{\rm p}$, $\nu_{\rm p}$ and
    $t_{\rm p}$.  The numbers are the power-law index to which the
    parameter in the relevant row is raised to give the scaling of the
    parameter in the selected column, such that, for instance, $S_{\rm
    p}\propto \nu_{\rm p}^{-0.93}$.  Correlations which a Spearman
    rank test show not to be significant at the 95 per cent level are
    shown in brackets.}}
\end{table}

There is clearly significant scatter in the individual events (due to
the strong dependences of $B$ and $\kappa$ on $S_{\rm p}$, $\nu_{\rm
p}$ and $\theta_{\rm src}$).  However, we can use equations
\ref{eq:magfield}, \ref{eq:enorm} and \ref{eq:selfabs} together with
the derived scalings of $S_{\rm p}$, $\nu_{\rm p}$ and $t_{\rm p}$
with one another, given in Table \ref{tab:scalings}, to show the
expected trends in $B$, $\kappa$ and $\tau$.  These are plotted as
dashed lines in Fig.~\ref{fig:physparms}.  Spearman rank correlations
show that the only significant trends are for the magnetic field to
increase with event duration (correlation coefficient 0.64, chance
probability $4\times10^{-4}$), the electron normalisation to increase
with event brightness (correlation coefficient 0.47, chance
probability 0.015) and decrease with peak frequency (correlation
coefficient 0.47, chance probability 0.034), and for the opacity to
increase with brightness and duration, and to decrease with frequency.
The correlations which are not found to be significant have the
shallowest power law indices (between $-1.0$ and $1.0$), suggesting
that they are the most easily masked by the scatter in the parameters.
The correlation of optical depth with event parameters is due to the
distance along the jet at which the peak occurs.  The shorter
outbursts peak at higher frequencies and closer to the core, and have
already reached their peak flux density at 15\,GHz by the time the jet
reaches a radius of 1\,AU, so they are optically thin.  The large,
bright outbursts have not yet expanded sufficiently to become
optically thin at 15\,GHz by the time the jet radius is 1\,AU.  The
calculated values for the physical parameters of the average outbursts
of 1994, 2001 and 2002 are listed in Table \ref{tab:physparms}.  
With the derived magnetic field strengths, the synchrotron loss
timescales at 43\,GHz range from 18\,d to 70\,y.  By the time the
shock has reached the radio photosphere, its size is not limited by
radiative losses, having instead been set earlier when the magnetic
field was higher, since when it has been determined by adiabatic
expansion.  Thus we cannot further constrain the size of the emitting
region using the radiative loss timescale.

\begin{table}
\begin{center}
\begin{tabular}{cccc}
\hline\hline
Parameter & 1994 & 2001 & 2002\\
\hline
$<S_{\rm p}>$ (Jy) & 1.4 & 9.3 & 0.14\\
$<\nu_{\rm p}>$ (GHz) & 4.5 & 3.6 & 26.9\\
$<t_{\rm p}>$ (h) & 16.8 & 31.9 & 0.47\\
$z_{\rm p}$ (AU) & 201 & 381 & 5.7\\
$R_{\rm p}$ (AU) & 7.0 & 13.3 & 0.20\\
$B_{\rm p}$ (G) & 0.25 & 0.03 & 0.13\\
$B(1 {\rm AU})$ (G) & 2.54 & 0.56 & 0.02\\
$\kappa_{\rm p}$ (erg$^{s-1}$cm$^{-3}$) &
$6.5\times10^{-2}$ & $1.9\times10^{0}$ & $2.1\times10^3$\\
$\kappa (1 {\rm AU})$  (erg$^{s-1}$cm$^{-3}$) & $1.4\times10^{1}$ & $2.3\times10^3$ & $2.5\times10^1$\\
$\tau_{\rm p}(15{\rm GHz})$ & $1.6\times10^{-3}$ & $8.2\times10^{-4}$ & $3.8\times10^{-1}$\\
$\tau(1 {\rm AU})(15{\rm GHz})$ & $5.6\times10^0$ & $4.3\times10^1$ & $4.3\times10^{-4}$\\
\hline
\end{tabular}
\end{center}
{\caption{\label{tab:physparms}A comparison of the fitted parameters
    of the different outbursts of Cygnus X-3 \citep[2002 January 25
    from this paper, and 2001 September and 1994 February-March
    from][]{Lin07}, and the physical quantities derived from them.
    For a given flaring sequence, the average time, frequency and flux
    density of the peak in the spectrum are $<S_{\rm p}>$, $<\nu_{\rm
    p}>$ and $<t_{\rm p}>$.  This occurs at the transition from the
    Compton to the adiabatic stage, a distance $z_{\rm p}$ downstream
    in the jet where the jet radius is $R_{\rm p}$.  At this point,
    the magnetic field is $B_{\rm p}(S_{\rm p},\nu_{\rm p},t_{\rm
    p})$, the electron normalisation is $\kappa_{\rm p}(S_{\rm
    p},\nu_{\rm p},t_{\rm p})$, and the optical depth to synchrotron
    self-absorption at 15\,GHz is $\tau_{\rm p}(S_{\rm p},\nu_{\rm
    p},t_{\rm p})$(15\,GHz).  Extrapolating these last three values
    back to the point where the jet radius is 1\,AU, for valid
    comparison between the flaring sequences, gives the magnetic field
    $B(1 {\rm AU})$, the electron normalisation $\kappa (1 {\rm AU})$
    and the optical depth $\tau(1 {\rm AU}$)(15\,GHz).}}
\end{table}

To investigate the decreasing opacity over the course of the flare
sequence suggested by \citet{Fen97} and seen in our analysis in
Section \ref{sec:opacity}, we performed a Spearman rank-order
correlation test between the start times of the individual outbursts
of 2002 January 25 and their peak flux densities, frequencies and
timescales shown in Table \ref{tab:shifts}.  The only significant
correlation was a trend for the peak flux density to decrease with
start time, with a Spearman rank coefficient of $-0.83$ ($<0.3$ per
cent probability of a chance correlation).  This trend, due to the
decay of the flare at the start of the observations, is seen in the
individual lightcurves of Fig.~\ref{fig:variability}.  There was no
significant correlation of optical depth with start time.  If the
opacity is indeed decreasing over the course of a flaring sequence,
that opacity is not generated by the synchrotron mechanism.  It could
be free-free absorption due to a decreasing proportion of entrained
thermal material \citep{Fen97}, although such a component was not
included in our modelling, so its effects cannot be quantified.  We
would need simultaneous well-sampled lightcurves at lower frequencies
to investigate its effects.

\subsection{The differences between large and small outbursts}
\label{sec:difference}

VLBI observations \citep{Mio01,Mil04,Tud07} have clearly demonstrated
that the outbursts of Cygnus X-3 vary in jet morphology as well as in
strength, showing both one-sided and two-sided jets, with the dominant
component moving to the south in some cases and to the north in
others.  Such differences cannot be explained by varying the magnetic
field strength or electron normalisation at the onset of the shocks.
Changes in the accretion flow, or in the immediate environment of the
binary system, possibly due to changes in the strength of the stellar
wind, are plausible explanations for such variations in the jet
morphology.  However, bearing such caveats in mind, we can attempt to
use our jet model fits to gain some insight into the differences
between large and small outbursts.  Varying the jet orientation and
environmental parameters will add scatter to the observed relations,
but we can hope to discern general trends.

We found that the same shock-in-jet model can explain the radio
lightcurves of Cygnus X-3 during giant outbursts, intermediate flares,
and low-level activity, suggesting that the three discrete
variability classes identified by \citet{Wal94} are similar phenomena,
which differ only in scale.  This indicates that a jet outflow
persists into the low-level active state in which Cygnus X-3 spends
the majority of its time, and that the formation of shocks is
responsible for the observed events in the radio lightcurves.
Outbursts which peak earlier do so at higher frequencies and with
lower peak flux densities.  The fitted model parameters strongly
suggest that the faster, weaker outbursts occur closer to the core,
whereas the brighter outbursts occur further downstream.  For a
constant jet speed (as assumed in our modelling), this implies that
shocks forming further downstream take longer to expand by a given
factor, giving rise to a longer timescale for the emission, and a
lower peak frequency.

While Fig.~\ref{fig:physparms} has considerable scatter, as expected
from the above discussion, the observed trends demonstrate that the
brighter, lower-frequency peaking outbursts have higher electron
normalisations.  The trends with magnetic field are not as clear-cut,
although the correlation of longer-duration outbursts with magnetic
field suggests that the brighter, lower-frequency peaking outbursts
should also have higher magnetic field strengths.  It is possible that
the shallowness of these correlations allow them to be masked by the
scatter in the parameters, making them appear less significant.  If
both the magnetic field and electron normalisation are higher for
stronger outbursts, it suggests that some significant fraction of the
magnetic field could be generated by the electrons themselves.
However, the dependences of the observable parameters $S_{\rm p}$,
$\nu_{\rm p}$ and $t_{\rm p}$ on the underlying physical parameters
$B$ and $\kappa$ are clearly complex, and better modelling is required
to disentangle the relative importance of the magnetic field and
electron normalisation in determining the size of the outbursts.

\subsection{Implications for high-resolution imaging}

While the lightcurves can be well fitted by shocks propagating
down a jet, it is natural to consider whether these events in fact
correspond to discrete VLBI components.  Hughes, Aller \& Aller (1989b,
1991) found that a model consisting of shocks propagating down a jet
could explain the observed evolution of the total intensity, polarized
intensity, and VLBI morphology of three extragalactic radio sources.

Several studies \citep[e.g.][]{Mut90, Abr96} have found that peaks in
total intensity lightcurves correspond to the ejection of new VLBI
components, with the flux density of the ejected components roughly
corresponding to the increase in total flux density.  Subsequently,
the decomposition of lightcurves into discrete flaring events
\citep{Val99, Tur99, Lin06} has found a good correspondence between
VLBI component zero-separation dates and the fitted start times of
those events. However, the component flux densities were less well
fitted, probably due to an underestimate by the model of the
contribution of the underlying jet.  The largest sample studied to
date, incorporating 27 blazars, found that every resolved VLBI
component had a corresponding flare in the total intensity lightcurve,
with zero-separation dates which corresponded well to the start dates
of the flares inferred from the lightcurves \citep{Sav02}.  But again,
while the flux densities of the flares correlated with those of the
resolved VLBI components, there was significant scatter about this
relation.

Flares in the total flux density are also observed which do not appear
to correspond to resolved ejecta on VLBI scales
\citep[e.g.][]{Pya06,Lin06}.  The favoured explanation
\citep[reinforced by evidence for unresolved components derived from a
closure phase analysis performed by][]{Sav02} is that such events do
produce ejecta, but that these have faded before propagating out to
the point where they can be resolved.  \citet{Sav02} therefore
suggested that every large ($>30$ per cent increase in flux density)
flare in the total intensity lightcurve corresponded to the formation
of a new shock.  This raises the question of what sets the decay time
of the ejecta.  If the shock takes a long time to become optically
thin at the observing frequency, the emission will have faded before
it can propagate out far enough to be resolved.  Similarly, if there
are significant radiative losses (particularly important at higher
frequencies), the emission will rapidly fade below the detection
limit.  In our model, the location of the outburst in the
three-dimensional frequency-time-flux density space will then
determine whether or not a given flare fades too rapidly to be
resolved at the observing frequency.  We note however, that the model
is an oversimplification, and that other events such as instabilities
in the flow, magnetic field enhancements, or interaction with the
surroundings could cause a rebrightening further downstream, allowing
components that would not otherwise be resolved to be detected.  While
we cannot hope to account for such phenomena without introducing many
extra free parameters while attempting to model the observed
lightcurves, they could help to account for the discrepancies between
model predictions and VLBI observations.

In the case of Cygnus X-3, VLBI jets have been seen out to angular
distances of 50--100\,mas \citep{Mio01,Mil04,Tud07}.  Since the peak
frequency $\nu_{\rm p}$ scales with distance (i.e.\ time) along the
jet as $z_{\rm p}^{-0.47}$ (Table \ref{tab:scalings}), the low-level
flares seen here, with peak frequencies ten times lower than those
seen in the giant flare of 2001, might be expected to be extended at
the 10--20\,mas level, assuming that the speeds of propagation were
similar in both cases.  However, the resolved knots seen during giant
flares were observed well after the peak of the outbursts, while the
emission was fading.  For an outburst peaking at flux densities 100
times lower, it would be challenging to detect extension at such low
levels.  Furthermore, travel-time arguments suggest that it would take
several hours for shocks travelling at $c$ to reach angular distances
of 10--20\,mas, for a source distance of 10\,kpc.  Indeed, the
constraints on the source sizes derived in Section \ref{sec:opacity}
were 0.2--0.3\,mas, smaller than the scattering disk and at the limit
of what can be achieved with current VLBI arrays, even at the highest
frequencies.  We therefore find it unlikely that these flaring events
could be resolved with current instrumentation.

\section{Absorption mechanisms}

In Section \ref{sec:opacity}, we quantified the opacity effects which
led to the delay and smoothing of the 15-GHz lightcurve with respect
to that at 43\,GHz.  However, the nature of such opacity was not
explicitly determined.  Plausible absorption mechanisms are
synchrotron self-absorption \citep{Mir98}, and free-free absorption
from either the wind of the companion star
\citep[e.g.][]{Fen95,Wal96}, or from entrained thermal electrons in
the jet \citep{Fen97}.

We can quantify the effect of the dense wind of the Wolf-Rayet
companion star, which causes delays as the emitting regions move out
from behind the radio photospheres, located closer to the central
binary at higher frequencies.  Using the parameters suggested by
\citet{Wal96} for the wind speed, temperature, ionisation parameters
and mass-loss rates would predict the 15 and 43\,GHz photospheres to
lie at $\sim4\times10^{13}$ and $\sim2\times10^{13}$\,cm respectively,
implying a delay of $\sim11$\,min between the two frequencies,
assuming propagation at the speed of light.  The speed of the ejecta
has been suggested to lie between $0.3c$ \citep{Sch95} and $0.81c$
\citep{Mio01}, corresponding to delays of 14--36\,min.  With only two
frequencies, we cannot verify whether the relation holds across all
frequency bands, but it is consistent with the observed dual-frequency
delay only if the speed of the ejecta is at the upper end of the
allowed range.  If this were the only source of opacity, the
decreasing delay seen over the course of the outburst sequence would
suggest variations in either the speed of the ejecta, or the location
of the photospheres between outbursts.  Since variations in the wind
would take $\geq12$\,h to propagate out to the photospheres at a wind
speed of $2000$\,km\,s$^{-1}$, we consider the latter possibility
unlikely.

The frequency-dependent delays found by \citet{Mol84} are inconsistent
with the predictions of free-free absorption in a wind.  The ratio of
delays between different pairs of frequencies can be compared with
that expected for emitting knots moving at constant speed between
photospheres whose radii scale as $\nu^{-0.67}$ \citep{Wri75}.  The
measured delay between emission at 2 and 20\,cm is longer than
predicted, suggesting the presence of an additional source of opacity
affecting the 20\,cm emission, even in small flares of amplitudes
comparable to those presented in this work.  As found by
\citet{Wal96}, while wind opacity may play a role, a variable
injection rate of relativistic electrons into the jet dominates the
observed lightcurves.  If wind opacity is not a significant
contributor to the overall delay, the photospheric radii must be
smaller than calculated, implying that the wind mass loss rate is
lower than the value of $10^{-5}M_{\odot}$\,y$^{-1}$ assumed by
\citet{Wal96}.

The shock-in-jet model of Section \ref{sec:modelling} is able to fit
the lightcurves fairly well at both frequencies, and thus is able to
account for the observed degree of delay and smoothing.  It takes no
account of any free-free opacity, so synchrotron self-absorption alone
is certainly capable of generating the observed delay.  The same model
also reproduces the giant outbursts, in which the shocks form further
downstream in the jet, beyond the radio photospheres formed by the
wind of the companion star, when synchrotron self-absorption or
free-free absorption by entrained thermal material \citep{Fen97} is
believed to be responsible for the frequency-dependent delays.  With
only two relatively high radio frequencies to determine the delay, it
is not possible to constrain the absorption mechanism further with our
data.  While the decreasing delay with time is suggestive of entrained
thermal material, the decrease is not as marked as found by
\citet{Fen97} for larger flares.

\section{Conclusions}
\label{sec:conclusions}
We have observed an episode of low-level activity in the X-ray binary
system Cygnus X-3, demonstrating that the source remains active in its
so-called quiescent state.  The 15-GHz lightcurve lags the 43-GHz
lightcurve by $9.5\pm1.0$\,min, and the variability is smoothed out at
the lower frequency.  There is no significant power on timescales
shorter than $\sim20$\,min at 43\,GHz, suggesting an approximate size
scale of $\sim2$\,AU for the emitting region at this frequency.  As
expected, the source was not seen to be extended at either frequency.
Linear polarization was detected at levels of 1--3.5 per cent at
43\,GHz, but not at 15\,GHz, suggesting that either opacity, beam or
line-of-sight Faraday depolarization is affecting the lower frequency.
The polarized flux density varies with the total intensity, with an
increase in polarized flux and a change in the electric vector
position angle during a flare in total intensity, consistent with the
formation of shocks in the flow aligning the magnetic field and giving
rise to enhanced emission at the shock front.

The lightcurves could be well fitted at all frequencies with the same
shock-in-jet model as has been applied to larger outbursts, suggesting
that the same phenomena are at work in causing these low-level flares
as are responsible for the giant outbursts.  Stronger flares are seen
to peak at lower frequencies and on longer timescales than fainter
outbursts, consistent with the formation of shocks further downstream
in the jet.  Brighter outbursts which peak at lower frequencies tend
to have higher electron normalisations, and we find evidence that the
strength of the magnetic field may also be higher in such outbursts.
However, the scatter in these relations demonstrates that the true
dependencies of the observables on the fundamental physical parameters
are significantly more complex than a simple power-law relation.
An in-depth investigation of a plasmon model is beyond the scope of
this paper, and on the basis of these data alone we cannot
definitively rule out a plasmon model in favour of a shock-in-jet
scenario.  However, the similarities we found to the larger outbursts
with better infrared coverage (a frequency regime where the
predictions of the two models can in principle be distinguished)
provide qualitative support for our application of the shock-in-jet
model to explain these small-scale flaring events.

In agreement with previous work, we find that free-free absorption in
the stellar wind of the companion star is unlikely to be responsible
for the observed opacity effects and the time delay between the two
frequencies.  The decreasing delay with time is suggestive of a
decreasing fraction of entrained thermal material, but with only two
frequencies, we cannot quantify the relative contributions of
synchrotron self-absorption and free-free absorption from entrained
thermal material.

\section*{Acknowledgments}
We would like to thank the referee, Philip Hughes, for his
constructive and insightful suggestions.  JCAM-J is a Jansky Fellow
of the National Radio Astronomy Observatory.  The NRAO is a facility
of the National Science Foundation operated under cooperative
agreement by Associated Universities, Inc.  ParselTongue was
developed in the context of the ALBUS project, which has benefited
from research funding from the European Community's sixth Framework
Programme under RadioNet R113CT 2003 5058187.  JCAM-J thanks
the UK Particle Physics and Astronomy Research Council and the
University of Oxford for support via a research studentship while
part of this work was carried out.
\label{lastpage}

\end{document}